# Electron mobility in $In_xGa_{1-x}N$ ($0.1 \leq x \leq 0.4$) channel HEMTs

Vikash K. Singh[1,2], Digbijoy N. Nath[,2,a)]

[1]Solid State Physics Laboratory, New Delhi 110054
[2]Centre for Nano Science and Engineering
Indian Institute of Science, Bangalore 560012 (India)

**Abstract**:

In this letter, we report on the theoretical investigations of electron mobility in *practically viable* designs of $In_xGa_{1-x}N$ (0.1<x<0.4) channel high electron mobility transistors (HEMT). Carriers in such devices are expected to exhibit a higher velocity and hence higher cut-off frequencies ($f_T$) for highly scaled architectures. We estimate that the mobility of two dimensional electron gas (2DEG) is limited by alloy scattering rather than phonon scattering unlike in conventional GaN-channel HEMTs. For x > 0.30, the mobility and sheet resistance are found to be < 500 cm$^2$/Vs and > 700 Ω/□ respectively, which can severely affect the parasitic voltage drop in access regions. The results presented here are believed to significantly guide the practical exploration of $In_xGa_{1-x}N$ channel HEMTs towards next-generation electronics by enabling careful design of device layouts in highly scaled transistors to minimize parasitic access region voltage drop which results due to significant degradation of 2DEG mobility.

---

a) Author to whom correspondence should be addressed.
   Electronic mail: digbijoy@cense.iisc.ernet.in

GaN based high electron mobility transistors (HEMT) with their superior material and electron transport properties, are being increasingly used in industrial and strategic sectors particularly in RF power amplifiers (eg: in radars) and high-power switching circuits (eg: in converters)[1]. However, they are nearing their high-frequency limits imposed by polar longitudinal (LO) optical phonons. The current highest record for cut-off frequency ($f_T$), for instance, is 450 GHz[2] for a 20-nm channel AlN/GaN/AlGaN HEMT. Although to a first order, the saturation velocity ($v_{sat}$) of electrons in conventional GaN-channel devices should correspond to the onset of LO phonon emission[3] at $v_{sat} = \sqrt{E_{op}/m_e} \approx 3 \times 10^7$ cm/s (where $E_{op}$ is LO phonon energy in GaN ~ 92 meV, $m_e$ is effective mass of electrons in GaN), yet measured electron drift velocity even in highly scaled RF GaN HEMTs is in the range of 1.5-2$\times 10^7$ cm/s[4,5]. This lowering of velocity is attributed[3] to the heating of electron gas at high fields due to inefficient heat transfer to the lattice by LO phonons. With highly-scaled self-aligned architectures along with regrown contact resistance reaching quantum limit[6], the contribution of parasitic delays in limiting the $f_T$ of GaN-channel HEMTs has been considerably reduced. Thus, to enable III-nitride Terahertz (THz) electronics, it is imperative to address the intrinsic delay which is dictated by the electron velocity in the channel as $\tau \sim L_{gate}/v_{sat}$. HEMTs with $In_xGa_{1-x}N$ as the channel are promising in this context.

The electron velocity in InN as well as in $In_xGa_{1-x}N$ is predicted[7,8,9] to be appreciably higher than that in GaN, particularly due to the lower effective mass in In-containing alloys. Further, under high-field conditions, the disorder in the alloy is expected to lead to more efficient heat transfer of non-equilibrium phonons to the lattice via acoustic phonons[10] which is therefore predicted to lead to an even more enhancement of velocity in $In_xGa_{1-x}N$ with x>0.2. Recently, using electro-absorption technique, the electron velocity in $In_{0.10}Ga_{0.90}N$ was measured to be 20% *higher* than in GaN at similar electric fields[11]. It therefore holds tremendous promise for velocity enhancement and hence for $f_T > 500$ GHz electronics with $In_xGa_{1-x}N$ (x>0.2) as channel in HEMTs. However, achieving InGaN-channel HEMT with high In-composition (x>0.10) has always presented epitaxial challenges in terms of material and surface qualities[12]. The electron mobility in such $In_xGa_{1-x}N$ (x>0.2) channel HEMTs is expected to be significantly degraded due to alloy scattering, but for a highly scaled device with self-aligned gate architecture, the role of electron velocity is expected to be much more predominant than that of the carrier mobility. There are a few reports on InGaN-channel (x<0.10) HEMTs[13,14,15], most of which explored InGaN as the channel not for a predicted velocity enhancement but for better carrier confinement with GaN buffer as back-barrier. Recently HEMT with $In_xGa_{1-x}N$ as a channel layer having composition in the range from x= 0.05 to 0.10 exhibiting experimentally measured electron mobility from 1070 to 1290 $cm^2$/V-s with associated sheet charge density of ~ 2 x $10^{13}$ $cm^{-2}$ have been demonstrated. Further, state-of-art DC and small-signal performance with $f_T/f_{max}$ = 260/220 GHz was also reported[16] in an $In_{0.05}Ga_{0.95}N$ channel HEMT.

Though the growing interest in $In_xGa_{1-x}N$ channel HEMTs has resulted in number of reports on study of temperature dependent electron mobility for a given composition in previous years, there has been no report to date which theoretically studies electron mobility in $In_xGa_{1-x}N$ channel devices with x>0.20 for *practically viable* HEMT designs[17,18,19,20]. The primary reason is the limitation imposed by composition dependent critical thickness of $In_xGa_{1-x}N$ on GaN which restricts the design of an epitaxial stack for x>0.20[21]. The electron mobility, though predicted to be of lower in significance than velocity in the intrinsic HEMT region, is expected to affect the parasitic access resistance by lowering the sheet resistance. Hence it is crucial to estimate the composition dependent 2DEG mobility in practically viable InGaN channel HEMT designs with x>0.20.

In this letter, we theoretically study the electron mobility as a function of alloy mole fraction in $In_xGa_{1-x}N$ (0.2<x<0.4) as well as its dependence on carrier concentration for Ga-polar HEMTs. We invoke 2D formalism[22] including contributions from polar LO phonon and alloy scattering mechanisms to estimate the mobility using Born approximation to Fermi Golden Rule[23]. We are assuming a smooth InGaN/GaN interface, and hence the role of interface roughness is excluded in the current study. The contributions from other scattering mechanisms such as acoustic phonon and impurity are expected to be minimal.

The schematic for epitaxial stack for InGaN channel HEMTs is shown in fig. 1(a). Composition and thickness of InGaN channel have been taken such to meet two considerations – firstly, InGaN should certainly be pseudomorphic on GaN. Secondly, as In-composition is increased, the piezoelectric polarization becomes higher and leads to a higher electric field in the channel[24]. This leads to the valence band approaching the Fermi level or even 'touching' it for some critical thickness of InGaN. This condition is avoided by our designs since it is established that under such circumstances[25], acceptor like states responsible for current collapse are likely to occur. Hence, our designs maintain an $In_xGa_{1-x}N$ channel thickness of 8nm to 3.5nm for x=0.1 to 0.4 while maintaining similar charge densities at equilibrium for device stacks with various In-compositions. $Al_{0.3}Ga_{0.7}N$ and AlN layers of thickness 10nm and 2nm respectively are used as barrier and interlayer to the channel. Bare surface of $Al_{0.3}Ga_{0.7}N$ is assumed to be pinned at 1.7eV at equilibrium by Ni/Au contacts typically used for gate. The energy band diagrams for the structures are simulated by using a 1-D Schrodinger-Poisson solver[26]. The band diagram for a device stack with 30% InGaN (5 nm) as channel is shown in Fig. 1(b). The equilibrium charge density in the HEMTs with various In-compositions is ~ $1.8 \times 10^{13}$ $cm^{-2}$.

To calculate momentum scattering rates and hence carrier mobility, we use a modified Fang Howard wavefunction[22] given by equation (1)

$$\psi(z) = M \exp(-k_b z) \qquad z < o$$
$$= N(z + z_0) \exp\left(-\frac{b}{z}\right) \qquad z > 0 \qquad (1)$$

Here $k_b$ and $b$ are variational parameters determined by minimization of energy[22]. In practice $k_b$ can be taken as equal to $2\sqrt{2m_b \Delta E_{C1}/\hbar^2}$, where $m_b$ is the electron effective mass in AlN barrier region and $\Delta E_{C1}$ is the conduction band discontinuity at AlN/ $In_xGa_{1-x}N$ interface (Fig. 2). Value of b is given by the term $\left[6m(x)e F_3/\hbar^2\right]^{1/3}$, $m(x)$ and $F_3$ are effective mass of electron and electric field in $In_xGa_{1-x}N$ channel. Constants $M$, $N$ and $z_0$ are determined by the continuity of $\psi(z)$ and $\left[\frac{1}{m(z)}\right]\left[\frac{d\psi(z)}{dz}\right]$ at the interface along with normalization of $\psi(z)$. The penetration of the wave-function in to the AlN barrier is found to have a probability of ~ $10^{-4}$. We should note that unlike in a ternary channel device, the alloy scattering in conventional AlGaN/GaN HEMT is due to the wave-function penetration in to the AlGaN barrier and *not* due to the binary GaN channel. The momentum scattering rates and electron mobility for 2DEG due to alloy scattering $(\mu_{alloy})$ and due to optical phonon scattering $(\mu_{op})$ can be evaluated from equations (2), (3) and (4) which are based on Born approximation to Fermi Golden Rule[22,24,27].

$$\mu_{alloy} = \frac{e}{m(x)^2} \frac{\hbar^3}{\Omega_0 [\delta V]^2 x(1-x)} \frac{1}{\int \psi^4(z)dz} \qquad (2)$$

$$\mu_{op} = \frac{\kappa^* \kappa_0 \hbar^2}{2\pi e \omega_0 m(x) N(T) G(k_0)} \left(1 + \frac{1-\exp(-y)}{y}\right) \qquad (3)$$

$$\frac{1}{\mu_{total}} = \frac{1}{\mu_{alloy}} + \frac{1}{\mu_{op}} \qquad (4)$$

Here, $\Omega_0$ is unit cell volume and $\delta V$ is the scattering potential and integral over the thickness of $In_xGa_{1-x}N$ channel is implied. In equation (3), $y = \pi\hbar^2 n_s / m(x)k_B T$, where $n_s$ is 2D electron gas (2DEG) concentration, $N(T)$ is the phonon number given by Bose-Einstein statistics. And, $1/\kappa^* = 1/\kappa_\infty - 1/\kappa_0$, $\kappa_\infty$ and $\kappa_0$ are high frequency and low frequency dielectric constants respectively in the InGaN channel. $k_0 = \sqrt{2m(x)\omega_0/\hbar}$, $\omega_0$ is optical phonon frequency and $G(k_0) = b(8b^2 + 9k_0 b + 3k_0^2)/8(k_0 + b)^3$ is the form factor[27]. The material parameters except for LO phonon energy are linearly interpolated between InN and GaN. For LO phonon energy in InGaN, a bowing factor is used[4,18,28,29]. To estimate the scattering, the only parameter required now is the field in the channel ($F_3$) as indicated in Fig. (2). Simple electrostatic analysis of the schematic band diagram (Fig. 2), gives the field $F_3$ as a function of the 2DEG concentration ($n_s$), where $\sigma_{P1}$ and $\sigma_{P2}$ are the total polarization charges at the AlGaN/AlN and AlN/InGaN interfaces respectively.

$$F_3 = A.n_s + B \qquad (5)$$

where the coefficients $A$ and $B$ are,

$$A = \left( \frac{e}{\varepsilon_s} - \frac{\pi\hbar^2}{em(x)} \right) \qquad (6)$$

$$B = \frac{1}{(t_1 - t_2)} \left[ (\Delta E_{C1} - \Delta E_{C0}) - \phi_B + \frac{1}{\varepsilon_S} \{\sigma_{P2}(t_1 - t_2) - \sigma_{P1} t_1\} \right] \qquad (7)$$

The conduction band discontinuities are taken as 65% of the difference of energy band gaps of the corresponding materials[4].

Fig. 3(a) and (b) show the estimated mobility for In-compositions of 20% and 40% in the InGaN channel HEMT due to alloy and LO phonon scattering as a function of 2DEG concentration at room temperature. It is found that, as expected, alloy scattering dominates over LO phonon scattering in limiting the mobility of electrons unlike in conventional AlGaN/GaN HEMT where room temperature mobility is mostly limited by phonon scattering. We also find that the 2DEG concentration has very little effect on alloy scattering and hence on total mobility. This is also highlighted in Fig. 4(a) where we plot the total electron mobility for various InGaN compositions as a function of 2DEG density. Fig. 4(b) shows the total mobility as a function of alloy composition for two different 2DEG densities of $10^{13}$ cm$^{-2}$ and $2 \times 10^{13}$ cm$^{-2}$. We observe that the mobility drops rapidly from ~ 1000 cm$^2$/Vs to 570 cm$^2$/Vs as In-composition is increased from 10% to 20% and thereafter decreases at a lesser rate. For instance, the mobility drops from 500 cm$^2$/Vs to 420 cm$^2$/Vs when the In-composition is increased from 30% to 40%. Inset to Fig. 4(b) shows the estimated sheet resistance ($R_{SH}$) for the same, indicating that $R_{SH}$ can reach as high as 700-800 Ω/□ for In-compositions of 30-40% in the channel. Experimentally measured values of mobility for InGaN channel HEMT which have been reported are also highlighted in Fig. 4(a) and are found to be in good agreement[9,15] with the results of our calculation. The slight underestimation of the mobility for 10% InGaN in our work compared to the results of ref. [9] can be attributed to various reasons. There is no report on measurement of the conduction band discontinuity between AlN and In$_x$Ga$_{1-x}$N which can be a source of uncertainty and hence deviation from experimental data in our calculations. Besides, linear interpolation of material parameters may not hold precisely true for InGaN. Also, ref. [9] uses a quaternary barrier while we adopt an AlGaN barrier, leading to slightly different electric field profiles. Finally, InGaN is known to exhibit compositional inhomogeneity over large areas in epitaxy[30,31]. Hence, the 10% InGaN channel reported in ref. [9] as estimated by X-ray diffraction, could most likely exhibit mild variations over the sample area and hence can potentially lead to a slightly higher extracted mobility. These concerns require further experimental studies to estimate mobility.

In conclusion, we estimated the electron mobility in *practically viable* $In_xGa_{1-x}N$ channel HEMTs for a wide range of In-compositions and found that alloy scattering is the dominant mechanism over LO phonon scattering unlike in conventional GaN-channel HEMTs. For x>0.30, the mobility and $R_{SH}$ are found to be < 500 cm$^2$/Vs and > 700 Ω/□ which can severely affect the parasitic voltage drop in access regions of even highly scaled devices. Although $In_xGa_{1-x}N$ channel HEMTs are touted to be promising for > 500 GHz high-frequency devices with a higher carrier velocity, yet the severe degradation of mobility and hence of sheet resistance for x>0.30 requires attention to careful design of device layout to exploit the advantages offered by a predicted higher velocity. The results presented here are believed to significantly guide the practical exploration of $In_xGa_{1-x}N$ channel HEMTs towards next-generation electronics.

**Figure captions:**

**Fig. 1(a):** Schematic of device epi-stack considered in this study

**Fig. 1(b):** Energy band diagram for a HEMT with 5 nm InGaN (30% In-mole fraction) obtained by using a 1-D Schrodinger Poisson solver

**Fig. 2:** Schematic of conduction band profile for a typical device structure to extract the 2DEG density-dependent field profile in the channel

**Fig. 3(a):** Estimated alloy and LO phonon scattering limited electron mobility in 5 nm InGaN (30% In-mole fraction) HEMT

**Fig. 3(b):** Estimated alloy and LO phonon scattering limited electron mobility in 3.5 nm InGaN (40% In-mole fraction) HEMT

**Fig. 4(a):** Estimated total electron mobility in InGaN channel HEMTs for various Indium compositions as a function of 2DEG density

**Fig. 4(b):** Estimated total mobility in InGaN channel HEMTs as a function of alloy composition. Inset: Alloy composition dependence of sheet resistance of 2DEG.

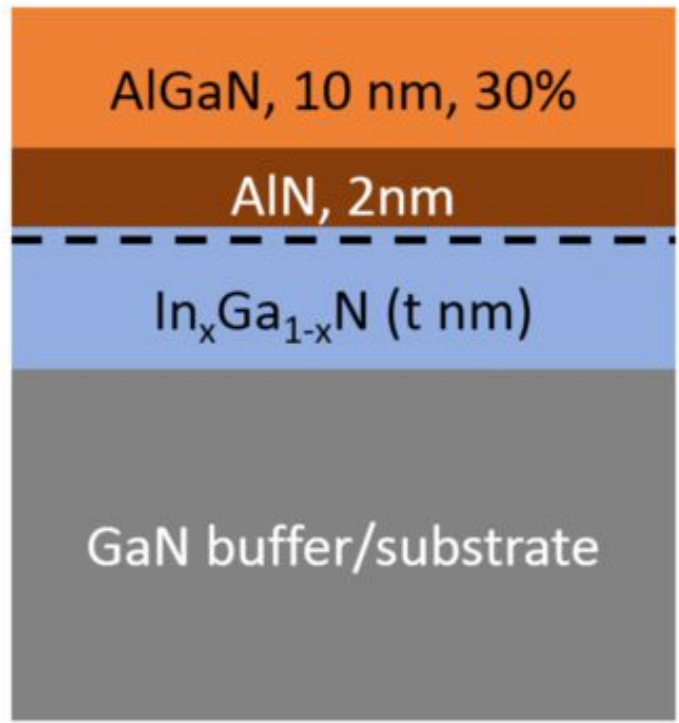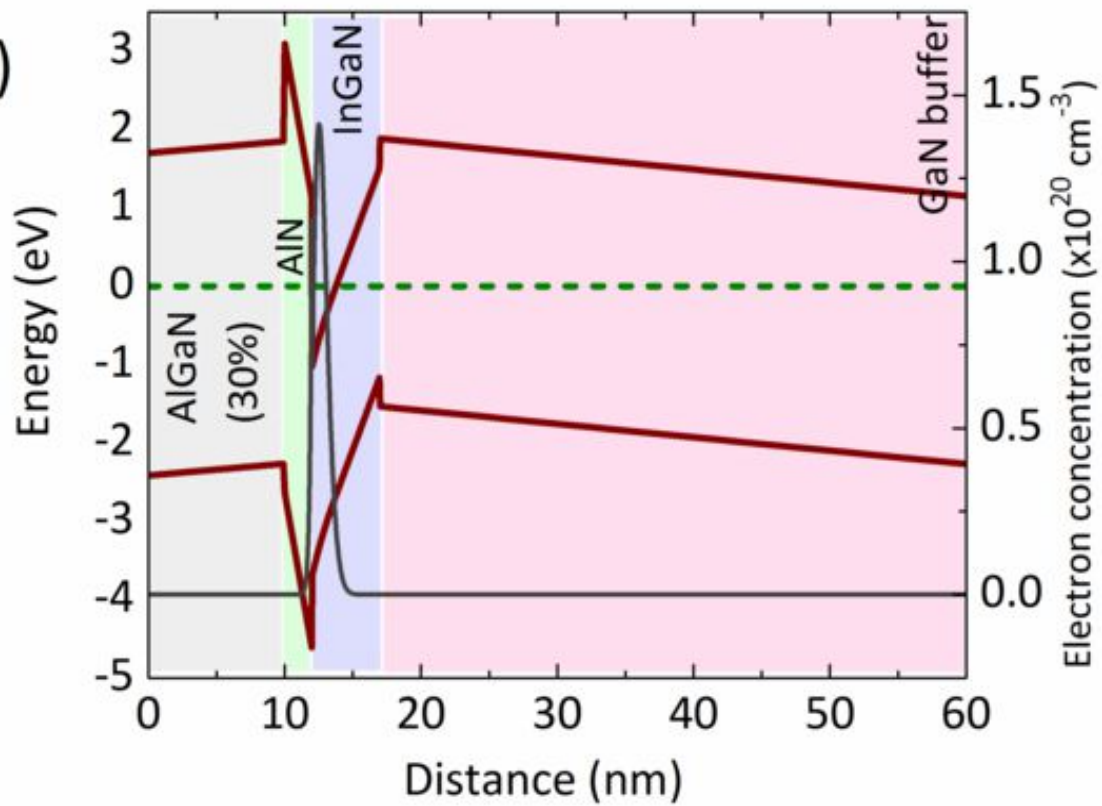

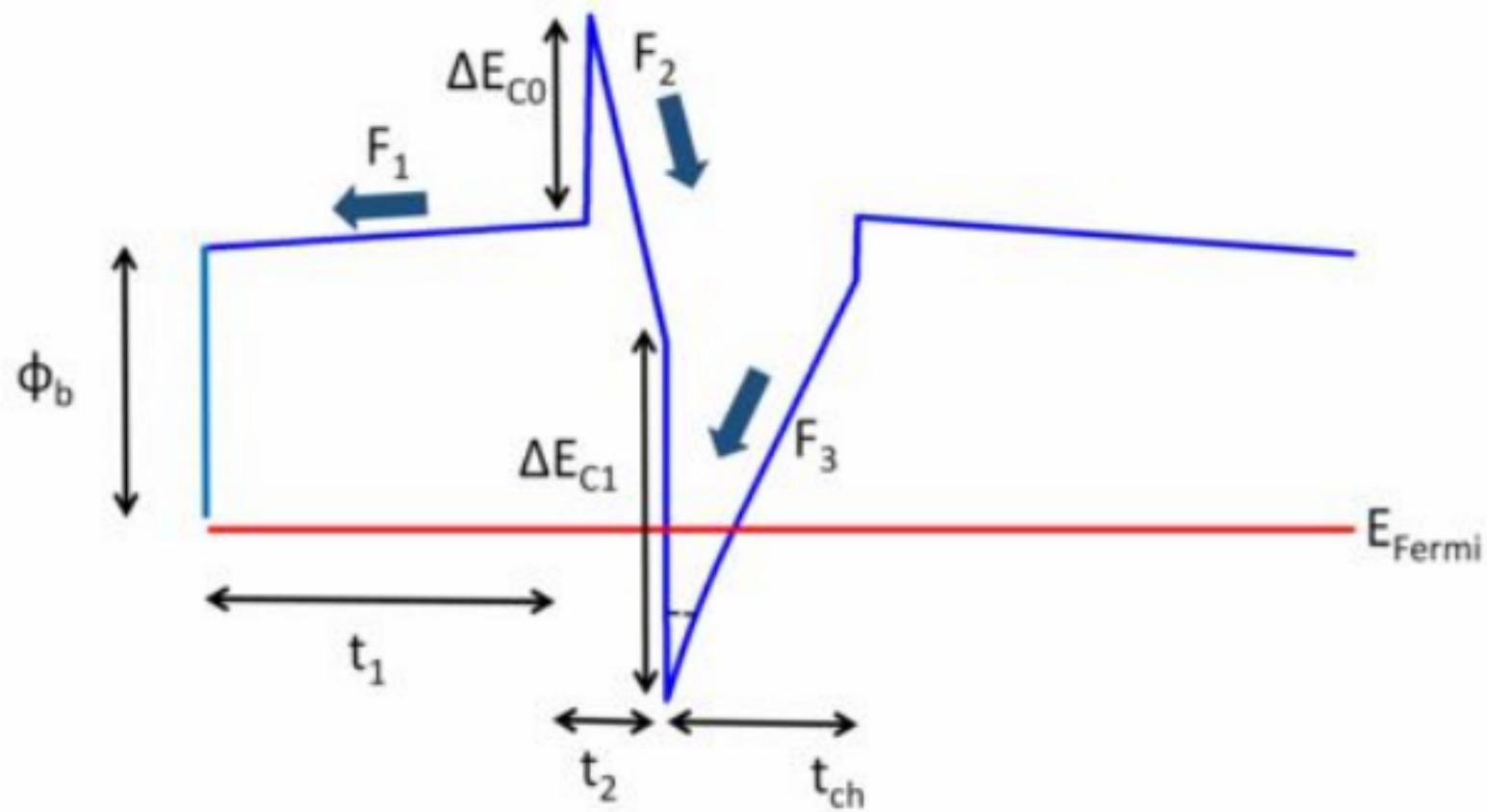

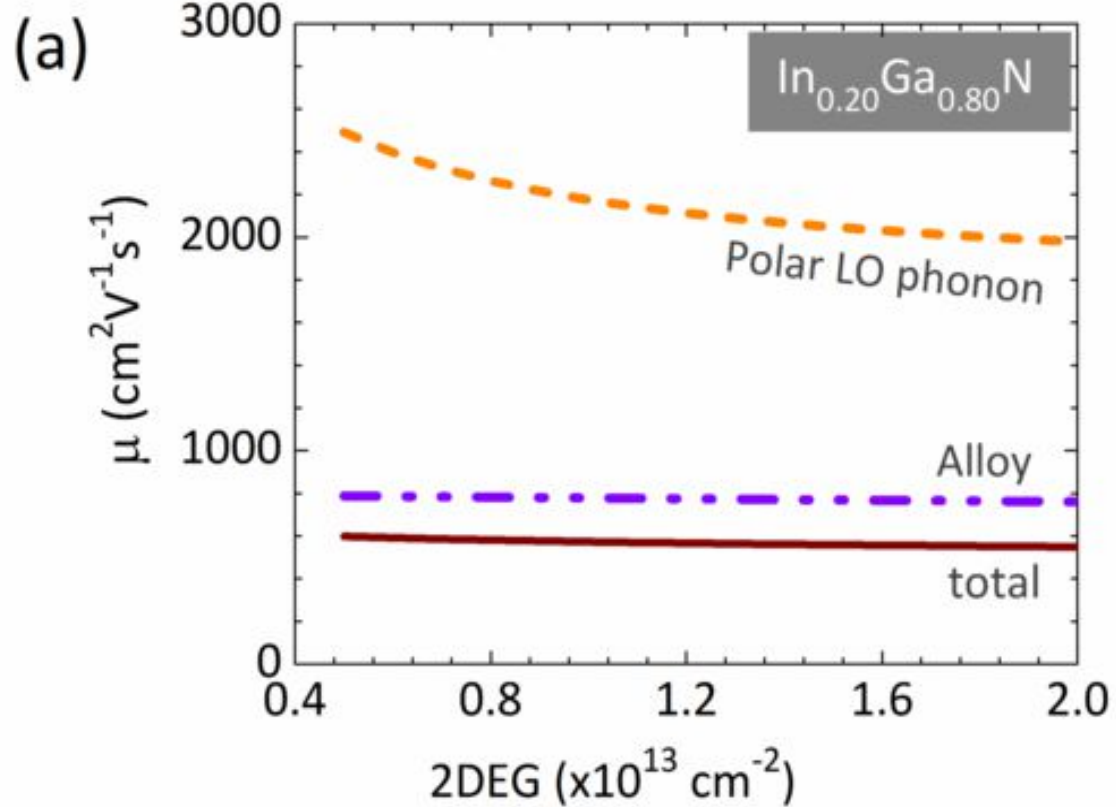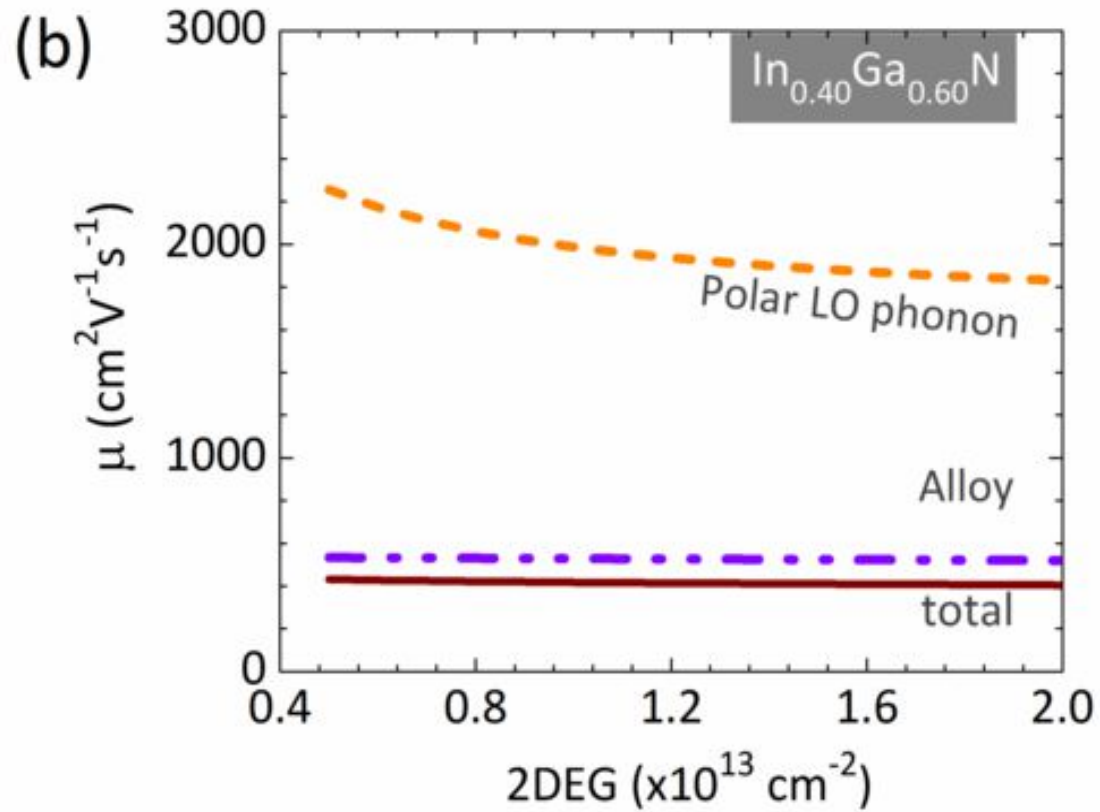

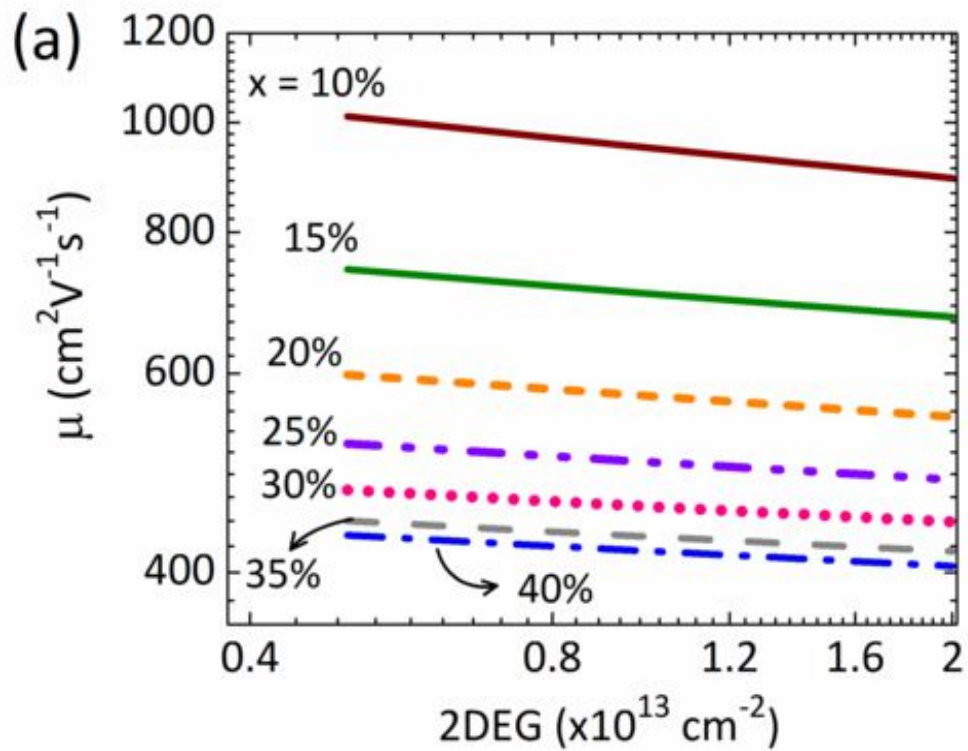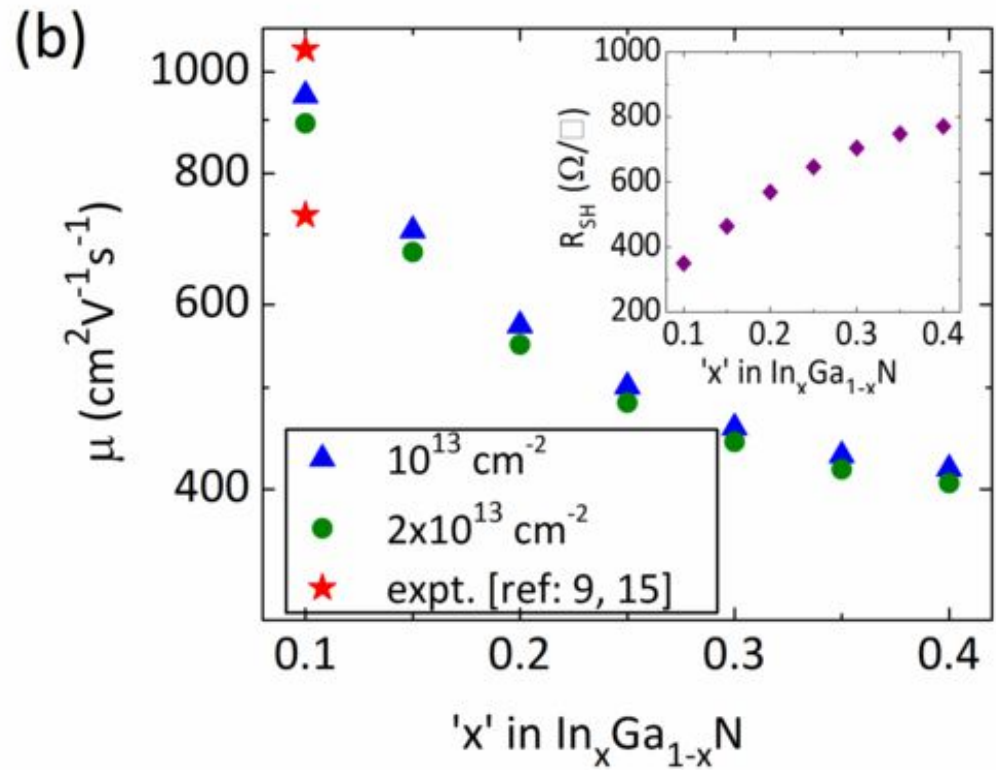